\documentclass[prb,reprint]{revtex4-1} 

\usepackage{amsmath}  
\usepackage{amsfonts} 
\usepackage{graphicx} 

\begin{document}


\title{Complex  Envelope  Variable Approximation in Nonlinear Dynamics.}

\author{Valeri V. Smirnov}
\email{vvs@polymer.chph.ras.ru} 
\altaffiliation[permanent address: ]{4 Kosygin Street, 
  Moscow, Russia} 
\affiliation{Department of Polymers and Composite Materials, Federal Research Center for Chemical Physics, RAS}

\author{Leonid I. Manevitch}
\email{manevitchleonid3@gmail.com}
\affiliation{Department of Polymers and Composite Materials, Federal Research Center for Chemical Physics, RAS}


\date{\today}

\begin{abstract}
We present the Complex Envelope Variable Approximation (CEVA) as the very useful and compact method for the analysis of the essentially nonlinear dynamical systems. 
It allows us to study both the stationary and non-stationary dynamics even in the cases, when any small parameters are absent in the initial problem.   
It is notable that the CEVA admits the analysis of the nonlinear normal modes and their resonant interactions in the discrete systems without any restrictions on the oscillation amplitudes. 
In this paper we formulate the CEVA's formalism and demonstrate some non-trivial examples of its application.  
The advantages of the method and possible problems are briefly discussed.
\end{abstract}

\maketitle 

\section{Introduction} \label{Int}
Modern level of  scientific researches and technologies  more frequently lead  to the problems, the core of which associates with the nonlinear dynamical processes. 
It concerns with the macro- as well as micro- and nano-scale phenomena.
If the first is the traditional scope of the nonlinear dynamics, the second is the new impetuous developed direction of the nonlinear studies. 
Generally speaking the nonlinear dynamics takes part in so large number of the sciences and technologies that the fact leads to formation of some inter-disciplinary "nonlinear science" \cite{Christiansen2000,Scott,Sagdeev88}.
However, in spite of that the nonlinear dynamics is required in everywhere, these problems remain very complex and are far from the completion.
Diversity and  complexity of the nonlinear dynamics lead  to the tedious and to frequently ambiguous approximations.
More one difficulty originates from that the numerical procedures do not allow to extrapolate the results obtained in one task into another one. 
This is one of the reasons why the interest to the problems, which would seem multiple studied, retain up to date.

There are many analytical and numerical methods, which cover the analysis of  the nonlinear problems  \cite{Bogoliubov1961,Krylov1947,Nayfeh1979,Sanders2007,Mickens2010,Esmailzadeh2019,Cveticanin2018,He2006,Leon1999}.  
The significant part of the most widely used approaches to the study of the nonlinear problems are the methods, which based on the asymptotic expansion of the solution into series of a small parameter.
In particular, the method of the multiscale expansion is based on the separation of the time scales, the ratio of which is determined by a small parameter \cite{Kevorkian1996,Leon1999}.
An elementary example can be found in the beating phenomenon in the system of the weakly coupled identical oscillators.
The process of the energy transfer from one oscillator to another one is determined by the frequency gap between in- and out-of-phase modes.
In such a formulation this problem nearly relates to the slowly varying envelope approximation, which has been considered by Van der Pol \cite{VanderPol_I,VanderPol_II}.
If the oscillators have the nonlinear characteristics, not only the slow energy transfer, but the energy capture on one of oscillators (energy localization) becomes possible \cite{Man07}.
Here one should emphasize that the description of the energy transfer and localization has been made in the terms of the complex representation of the variables \cite{Manevitch2001,Kosevich1989}, that is near analogue to the second quantization formalism \cite{Dirac,Haken}. 
 Such an approach in the combination with the multiscale expansion turns out to be very successful in the investigation of the wide class of the nonlinear problems: the coupled nonlinear oscillators \cite{Man07,CISM2010}, the forced nonlinear oscillators \cite{Musienko2009,CAP2016,AKovaleva2013}, the energy transfer and localization in the 1D nonlinear lattices \cite{Smirnov2010,CAP2016,Smirnov2017}, the mode coupling and energy localization in the carbon nanotubes \cite{Smirnov2016PhysD,Smirnov2018ND}, the synchronization of the self-excited oscillators \cite{MKovaleva2013} and the classic analogue of the superradiant quantum transition \cite{MKovaleva2013DAN}, the nonlinear passive control and energy sink \cite{Lamarque2020,Vakakis2014}, the problem of rotation stability of coupled pendulums \cite{Smirnov2019}.
 In a number of cases the complex representation of the variables allows us to find the stationary single-frequency solution (nonlinear normal modes) for the essentially nonlinear systems without any assumptions about oscillation amplitudes and without using any small parameter \cite{CAP2016_2,Smirnov2017}.
 Thus, the large advantage of this approach is that we obtain the main approximation, which satisfies to the initial essentially nonlinear problem.
 The non-stationary dynamics can be studied in terms of the slowly changed envelopes.  One should note that no restriction on the varying amplitude of the non-stationary oscillations arises, but the main requirement is a frequency closeness of the non-stationary and stationary solutions.
 In particular, such conditions allow us to study the nonlinear mode interactions in the discrete extended systems, if the lengths of the latter are large enough \cite{Smirnov2010,CAP2016_2,Smirnov2017}.
 In such a case the slow time scale is naturally appeared from the smallness of the inter-mode frequency gap.
 
 Generally speaking, the approach discussed below is a hybrid one.
 Actually, it  is formally similar to the Van der Pol slow varying envelope approximation \cite{VanderPol_I,VanderPol_II}, and, in some meaning, is close to the harmonic balance method \cite{Mickens2010}.
 On the other hand the multiple scale expansion is the essential constituent of the consideration of the non-stationary dynamics, but the post-factum revealing the small parameter makes it close to semi-inverse methods.
 Taking into account mentioned above  we will refer to the discussed approach as the "Complex Envelope Variable Approximation" (CEVA).
 The current work is aimed to describe the CEVA's formalism in a general case (section \ref{Method}) and to demonstrate some examples, in which the CEVA's advantages can be revealed (section \ref{Example}).
 The section \ref{Conclusion} contains the short discussion and conclusions.

\section{The Complex Envelope Variable Approximation}\label{Method}

Let us consider the nonlinear dynamical system the evolution of which is determined by the equation

\begin{eqnarray}\label{eq:system0}
\frac{d^2 u}{d t^2}+F(u)=0
\end{eqnarray}
where $u$ is the function of time $t$ and $F$ is a nonlinear function of $u$.
First of all we should note that we do not single out any parts, which contain a small parameter.
As it was emphasized in Introduction we want to study the slow processes which are  not  dictated by the external factors.
In this way, it is convenient to introduce the new dynamical variables:
\begin{eqnarray}\label{eq:Psi}
\Psi=\frac{1}{\sqrt{2}} \left(\sqrt{\omega} u + i \frac{1}{\sqrt{\omega}} \frac{d u}{d t} \right),
\end{eqnarray}
where $\omega$ is the frequency, which should be determined as the function of the oscillation amplitude.
The inverse transformation is well known:
\begin{eqnarray}\label{eq:inverse}
u = \frac{1}{\sqrt{2 \omega}} \left(\Psi+\Psi^{*} \right); \quad \frac{d u}{d t}= - i \sqrt{\frac{\omega}{2}} \left( \Psi - \Psi^* \right)
\end{eqnarray}
Using these variables one can rewrite equation (\ref{eq:system0}) as follows:
\begin{eqnarray}\label{eq:complexfull}
i \frac{d \Psi}{d t} - \frac{\omega}{2} \left(\Psi-\Psi^* \right) -\frac{1}{\sqrt{2\omega}} F\left(\frac{1}{\sqrt{2\omega}} \left( \Psi + \Psi^* \right) \right) =0
\end{eqnarray}
In the number of the nonlinear systems we  interest ourselves in the single-frequency stationary solution of equation (\ref{eq:system0}).
In order to get it we suppose next form of the solution:
\begin{eqnarray}\label{eq:Psistat}
\Psi = \psi e^{- i \omega t}.
\end{eqnarray}
This representation corresponds to the first term of the Fourier series and $\psi$ is assumed as a constant.
Expanding function $F$ into the Taylor series and averaging equation (\ref{eq:complexfull}) over the period $2 \pi /\omega$ allows us to extract the secular term.
Finally we get the equation for function $\psi$ in the form:
\begin{eqnarray}\label{eq:stateq0}
\frac{\omega}{2} \psi -\frac{1}{\sqrt{2 \omega}}\tilde{ \Phi} \left( \psi, \psi^*; \omega \right) =0.
\end{eqnarray}
At first glance equation (\ref{eq:stateq0}) is not more simple than the initial one.
However, for the number of the actual systems function $\tilde{\Phi}$ has the noteworthy structure:
\begin{eqnarray}\label{eq:Phi0}
\tilde{\Phi}\left(\psi,\psi^*;\omega\right) = \sum_{k=0}^{\infty}{c_k \left(\sqrt{\frac{2}{\omega}}\right)^{2k+1}|\psi|^{2k}} \psi
\end{eqnarray}
As it is shown in Appendix, some "good" nonlinearities admit the representation of the infinite sum in equation (\ref{eq:Phi0}) in the term of the special functions.
Equation (\ref{eq:stateq0}) should be considered as the amplitude-frequency relation, taking into account the relationship between complex value $\psi$ and the oscillation amplitude.
The latter follows from the first of equations (\ref{eq:inverse}) and expression (\ref{eq:Psistat}).
If $A$ is the oscillation amplitude, the modulus of value $\psi$ can be written as follows:
\begin{eqnarray}\label{eq:Ypsi}
|\psi | = \sqrt{\frac{\omega}{2}} A
\end{eqnarray}
In such a case, equation (\ref{eq:stateq0}) should be written in the form:
\begin{eqnarray}\label{eq:stateq1}
\frac{\omega^2}{2} A - \Phi(A)=0.
\end{eqnarray}
This equation allows us to find  oscillation frequency $\omega$:
\begin{eqnarray}\label{eq:freefreq}
\omega=\sqrt{\frac{2}{A} \Phi(A)}
\end{eqnarray}
This relation exhausts the stationary problem of the free oscillations of the system with one degree of freedom.

In order to illustrate this procedure efficiency, we find the amplitude-frequency relation for the most known nonlinear system - the pendulum.
In such a case function $F$ in equation (\ref{eq:system0}) is $\sin{u}$.
It was shown early \cite{CAP2016} the respective "secular" term in equation (\ref{eq:stateq0}) is read as follows
\begin{eqnarray}\label{eq:pendulumPhi}
\frac{\omega}{2} \psi -\frac{1}{\sqrt{2 \omega}} J_1\left(\sqrt{\frac{2}{\omega}} |\psi | \right) \frac{\psi}{| \psi |}=0,
\end{eqnarray}
where $J_1$ is the Bessel function of first order.
It is easy to see that according to relation (\ref{eq:Ypsi}), the argument of the Bessel function is the oscillation amplitude $A$.
Therefore, expression (\ref{eq:freefreq}) transforms into form
\begin{eqnarray}\label{eq:pendulumfreq}
\omega=\sqrt{\frac{2}{A}J_1(A)}.
\end{eqnarray}
Figure 1 shows the comparison of frequency 
 (\ref{eq:pendulumfreq}) with the exact value $\omega=\pi /2 K(\sin^2{(A/2)})$ (K is the full elliptic integral of first kind).
\begin{figure}
\centering{
\includegraphics[scale=0.2]{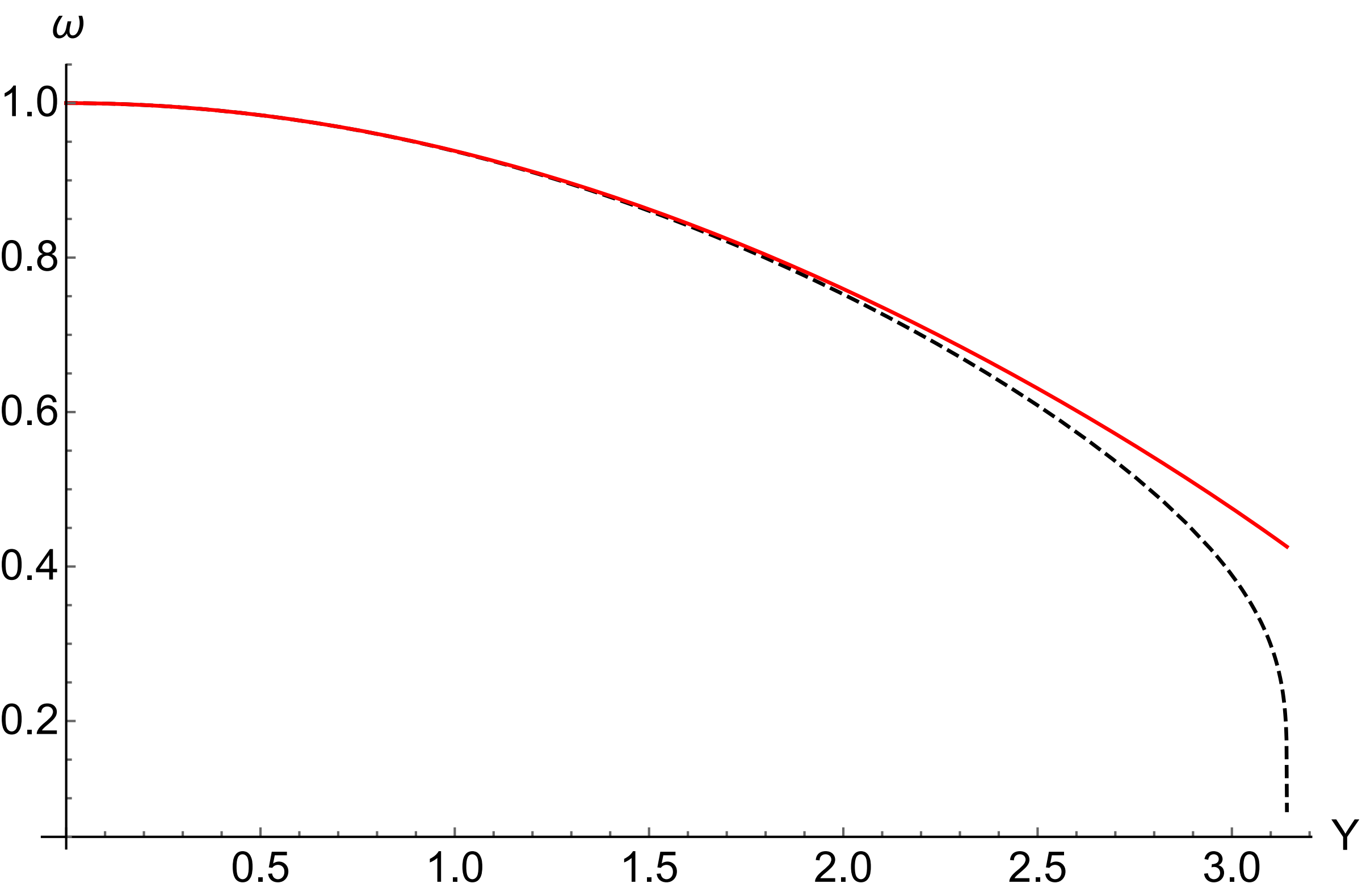}
\caption{The amplitude-frequency dependence in accordance to equation (\ref{eq:pendulumfreq}) (red solid curve) and exact value $\omega_0$ (black dashed curve).}
}
\label{fig:pendulumfreq}
\end{figure}
Analysing figure 1 one can conclude that approximation (\ref{eq:pendulumfreq}) is well enough in the wide interval of the oscillation amplitudes excluding a vicinity of the limiting value $A=\pi$.
It is the expected result because the motion with the amplitude  $A=\pi$ corresponds to the separatrix and evidently does not belong to the category of the single-frequency solutions  (\ref{eq:Psistat}).

The stationary state with complex amplitude $\psi$ and corresponding frequency $\omega$ has the energy
\begin{eqnarray}\label{eq:Estat0}
E= \frac{\omega}{2} |\psi |^2 - G(\psi; \omega),
\end{eqnarray}
where function $\tilde{\Phi}$ is coupled with $G$ by the relation
\begin{eqnarray}
\tilde{\Phi}=\frac{\partial G}{\partial \psi^*}
\end{eqnarray}

Let us consider energy (\ref{eq:Estat0}) as the Hamilton function of the system, which is parametrized by frequency $\omega$:

\begin{eqnarray}\label{eq:H1}
H = \frac{\omega}{2} | \psi |^2 - G\left(\psi; \omega \right)
\end{eqnarray}
The equation of motion can be obtained as follows:
\begin{eqnarray}\label{eq:nonstateqn}
i \frac{d \psi}{d \tau} = - \frac{\partial H}{\partial \psi^*}.
\end{eqnarray}
The latter leads to the time-dependent version of equation (\ref{eq:stateq0}):
\begin{eqnarray}\label{eq:nonstateq0}
i \frac{d \psi}{d \tau}+\frac{\omega}{2} \psi - \frac{1}{\sqrt{2 \omega}} \tilde{\Phi} \left( \psi; \omega \right) = 0
\end{eqnarray}
It is just the right time now to ask the question: what is a time scale of  variable $\tau$?
First of all, function $\psi$ is the envelope for single-frequency oscillations $e^{-i \omega t}$, therefore, it should be slowly changing.
In opposite case the averaging made above turns out to be invalid.
So, we can only consider the non-stationary motions with frequencies, those  weakly distinct from the frequency of stationary solutions.
In such a sense, the development  of non-stationary equations is similar to the slowly varying envelope approximation \cite{VanderPol_I,VanderPol_II}, and the slowness is determined by the structure of the equations obtained.
We should consider a slow motion, but we are not bound by the values of the amplitude variation.
I.e., only the request is that the specific time of the changing amplitude should be essentially large than the oscillation period.

Let us introduce the polar representation of function $\psi$.
\begin{eqnarray}\label{eq:psipolar}
\psi = a e^{i \delta}
\end{eqnarray}
Variables $a^2$ and $\delta$ form the canonical set for Hamilton function (\ref{eq:H1}).
Therefore, the equations of motion in terms of polar variables should be represented in form:
\begin{eqnarray}\label{eq:adelta0}
\frac{d a}{d \tau} = - \frac{1}{2a}\frac{\partial H}{\partial \delta} ; \quad  \frac{d \delta}{d \tau} = \frac{1}{2 a} \frac{\partial H}{\partial a}
\end{eqnarray}

We use the pendulum in order to make certain that equations (\ref{eq:adelta0}) actually describe the slow-time evolution of the system under a small disturbance of the pendulum stationary oscillations.
  The Hamilton function of the pendulum can be written as follows:
  \begin{eqnarray}\label{eq:Hpendulum}
  H = \frac{\omega}{2} a^2 - J_0 \left(\sqrt{\frac{2}{\omega}} a \right).
  \end{eqnarray}
Let us $\psi = \left( a + \alpha \right) e^{i \delta}$, where $a$ is solution of equation (\ref{eq:stateq0}) and $\alpha \ll a$ is a disturbance.  
  Taking into account equations (\ref{eq:adelta0}), we get
  \begin{eqnarray}\label{eq:pendulumdist} 
    \frac{d \alpha}{d \tau}  =0 \qquad \qquad \qquad \qquad \qquad \qquad \quad \\ \nonumber 
  \frac{d \delta}{d \tau}  \approx \frac{1}{a} \left( \frac{\omega}{2} a - \frac{1}{\sqrt{2 \omega}} J_1 \left( \sqrt{\frac{2}{\omega}} a \right) \right) + \\ \nonumber 
  \frac{1}{\omega a} J_2 \left( \sqrt{\frac{2}{\omega}} a \right) \alpha   
  +\left(\frac{J_1\left(\sqrt{\frac{2}{\omega}} a \right)}{\sqrt{2} a \omega ^{3/2}}-\frac{3 J_2\left(\sqrt{\frac{2}{\omega}} a \right)}{2 a^2 \omega }\right) \alpha ^2 
  \end{eqnarray}
 The first of equations (\ref{eq:pendulumdist}) shows that the amplitude of the disturbed motion is not change.
 The first term in the right hand side of equations (\ref{eq:pendulumdist}) is equal to zero and varying of phase $\delta$ turn out to be proportional to $\Delta \omega \sim \alpha d \omega /d a $.
 Due to that $\alpha=const$, it means that the disturbed solution runs off with constant velocity from the initial solution: $\delta \sim \Delta \omega \, \tau$.
 One should emphasize that the time scale of variable $\tau$ is controlled by the smallness of the right hand side of equations (\ref{eq:pendulumdist}).
 The latter can be determined by the smallness of either disturbance's amplitude ($\alpha \ll a$ ) or  smoothness of the oscillation frequency ($d \omega /d a \ll \omega/a$).
 
As a conclusion of this section, let us check that equations (\ref{eq:adelta0}) correctly predict the frequency changing in the limit of small amplitude $a$.
Assuming $a \rightarrow 0$ and $\omega \rightarrow 1$, and taking into account the relations between modulus of complex function and the oscillations' amplitude, we get
\begin{eqnarray}\label{eq:frequencycorrection}
\frac{d \delta}{d \tau} \approx \frac{\alpha^2}{8} =\frac{A^2}{16}.
\end{eqnarray}
 The sign of the correction is positive because solution (\ref{eq:Psistat}) contains $e^{-i \omega t}$.
 Correction (\ref{eq:frequencycorrection}) accords with  the expansion of exact pendulum frequency $\omega \approx 1- A^2/16$.
 
 In the next section we consider some examples and generalizations of the procedures discussed above.
 Before do it, let us formulate the standard steps for development of the evolution equations in the framework of the CEVA.
  
 i) Introduce of the complex variables (\ref{eq:Psi}).
 
 ii) Rewrite initial equation (\ref{eq:system0}) in terms of complex variables.
 
 iii) Extract the secular term for the single-frequency solution (\ref{eq:Psistat})
 
 iv) Use relation (\ref{eq:Ypsi}) in order to determine the stationary amplitude-frequency relation.
 
 v) Find the Hamilton function corresponding to the stationary solution.
 
 vi) Develop the non-stationary equations in accordance with the relations (\ref{eq:nonstateqn}) or  (\ref{eq:adelta0}).
 
\section{Examples and Applications}\label{Example}
\subsection{Forced damped oscillation of pendulum}\label{FDP}
The effect of the external forcing and dissipative processes often takes an important role in the dynamics of the nonlinear  systems.
 Therefore, we start this sections considering the  forced oscillations of the pendulum with the viscous friction.
Let us suppose that the pendulum undergoes the effect of external field $F(t)=f \cos{\omega t}$.
The stationary equation of motion can be written in terms of complex variable $\psi$ as follows:
\begin{eqnarray}\label{eq:FDpendulum}
\frac{\omega}{2} \psi - \frac{1}{\sqrt{2 \omega}} J_1 \left( \sqrt{\frac{2}{\omega}} | \psi | \right) \frac{\psi}{| \psi |} + i \frac{\nu}{2} \psi = - \frac{f}{2 \sqrt{2 \omega}},
\end{eqnarray}
where $\nu$ is the  coefficient of the viscous friction.
Using relation (\ref{eq:Ypsi}), it is easy to find the amplitude-frequency relation for the non-dissipative system ($\nu=0$):
\begin{eqnarray}\label{eq:forcedfrequency}
\omega^2=\frac{2}{A} \left( J_1 \left(A \right) -f \right).
\end{eqnarray}
However, function $\psi$ becomes complex for non-zero friction.
Assuming $\psi=x+i y$, we should  separate the real and imagine parts of equation (\ref{eq:FDpendulum}).
After some manipulations we can write the amplitude-frequency relation in the form:
\begin{eqnarray}\label{eq:FDAFR}
\frac{2}{\omega} \left(x^2+y^2 \right)=A^2=\frac{f^2}{\nu ^2 \omega ^2+\left(\omega ^2-\Omega ^2\right)^2} \\ \nonumber
\frac{y}{x}=\tan{\delta}=-\frac{\nu  \omega }{\omega ^2-\Omega ^2},
\end{eqnarray}
where $\Omega=\Omega(A)=\sqrt{2 J_1(A)/A}$ is the frequency of free oscillations with the amplitude $A$.
One should note that the first equation of (\ref{eq:FDAFR}) is the transcendent equation with respect to amplitude $A$.
It can be solved numerically and the result is represented in figure 2.
It is noteworthy that the amplitude-frequency relation (\ref{eq:FDAFR}) looks absolutely similar to its linear analogue, with the difference that the frequency of non-linear free oscillations  with amplitude $A$ plays the role of own frequency of the linear oscillator.
The expression for phase shift $\delta$ also has the same form as for the linear system.
\begin{figure}
\centering{
\includegraphics[scale=0.3]{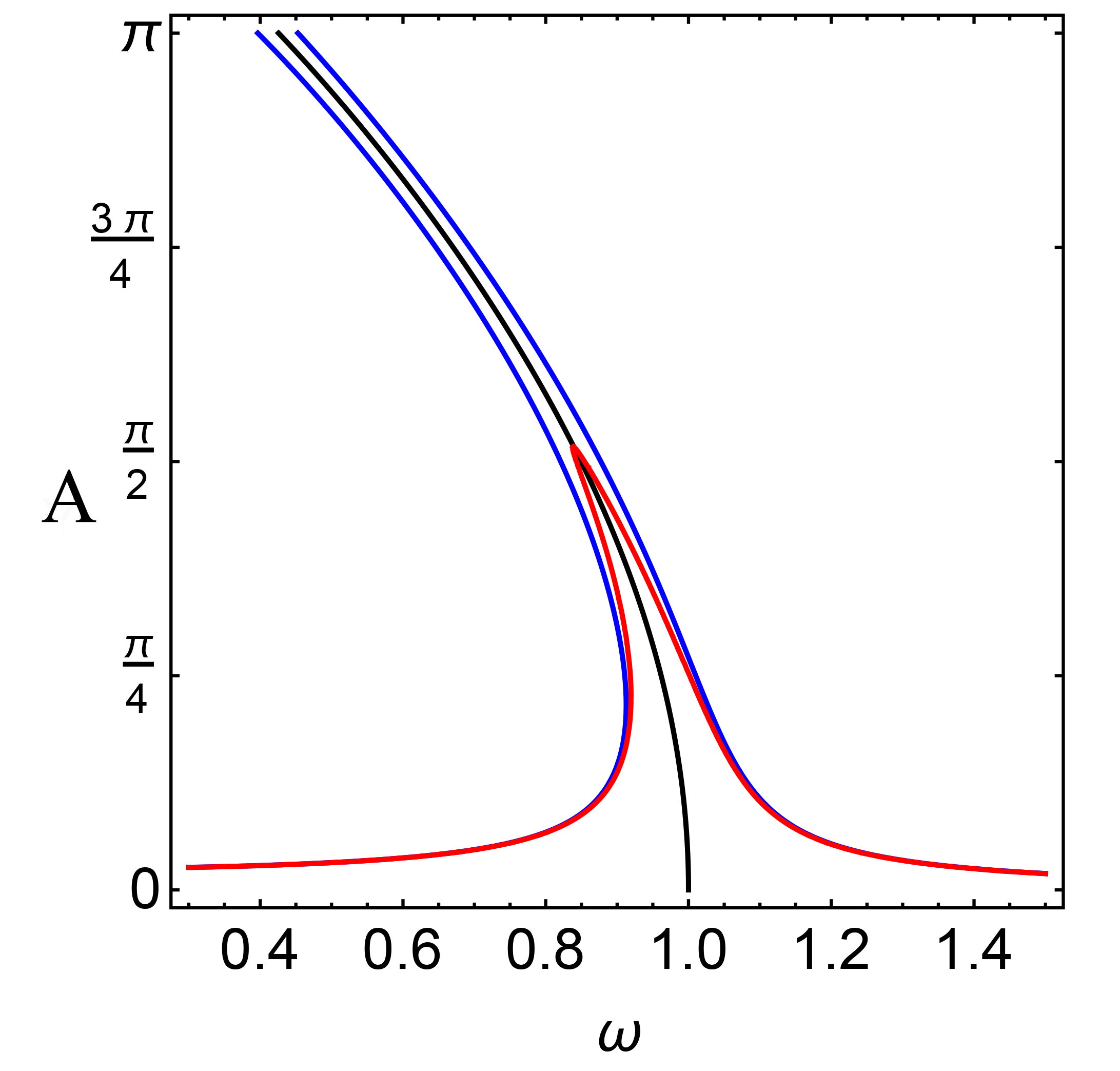}
\caption{Amplitude-frequency relations (\ref{eq:pendulumfreq}, \ref{eq:forcedfrequency}, \ref{eq:FDAFR})  for the free and forced oscillation of the pendulum without and with friction (black, blue  and red curves, respectively). Amplitude of the external force $f=0.075$ and friction $\nu=0.055$.  }
}
\label{fig:pendulumAFR}
\end{figure}

\subsection{Escape from a potential well}\label{Escape}
We illustrate the non-stationary nonlinear dynamics  considering the escape from potential well  under effect of single-frequency external field $F=f \cos{\omega\,t}$ without friction.
The transition of the pendulum from oscillations to the rotation gives the very clear example of such processes.
Writing function $\psi$ in polar form (\ref{eq:psipolar}) one can ascertain that the energy of forced oscillation is represented as follows:
\begin{eqnarray}\label{eq:Hforced}
H_f=\frac{\omega}{2} a^2 - J_0 \left( \sqrt{\frac{2}{\omega}} a \right) +\frac{a f \cos{\delta}}{\sqrt{2 \omega}}
\end{eqnarray}
We would like to find what combinations of the force's frequency $\omega$ and amplitude $f$ lead to the rotation of the pendulum, if the initial conditions are zero.
Such a problem is non-trivial even in the case of the simple parabolic potential \cite{Gendelman2019}.
In order to determine the boundaries of oscillations in the plane "frequency - force amplitude" one should analyse the possible non-stationary trajectories corresponding to Hamilton form (\ref{eq:Hforced}).
Let us consider the phase space in the terms $\{a, \delta \}$ assuming the frequency $\omega$ and force amplitude $f$ as the parameters.
Because hamiltonian (\ref{eq:Hforced}) does not contain the variables, which depends on the "fast" time $t$, the stationary oscillations correspond to the stationary points on the phase plane $\{ \delta, a \}$.
There are three stationary states in the low-frequency region in Fig. 2  and only one stationary state occurs if the frequency is larger than some value $\omega_*$.
The latter can be found as the root of the equation:
\begin{eqnarray}\label{eq:bifurcation13}
\frac{d \omega}{d A}=\frac{f-A J_2(A)}{\sqrt{2} A^{3/2} \sqrt{J_1(A)-f}}=0
\end{eqnarray}  
 Figure 3 shows the phase portraits of the system (\ref{eq:Hforced}) with the different values of frequency $\omega$ and the constant value of the force amplitude $f$.
 \begin{figure*}
 \centering{
a \includegraphics[scale=0.15]{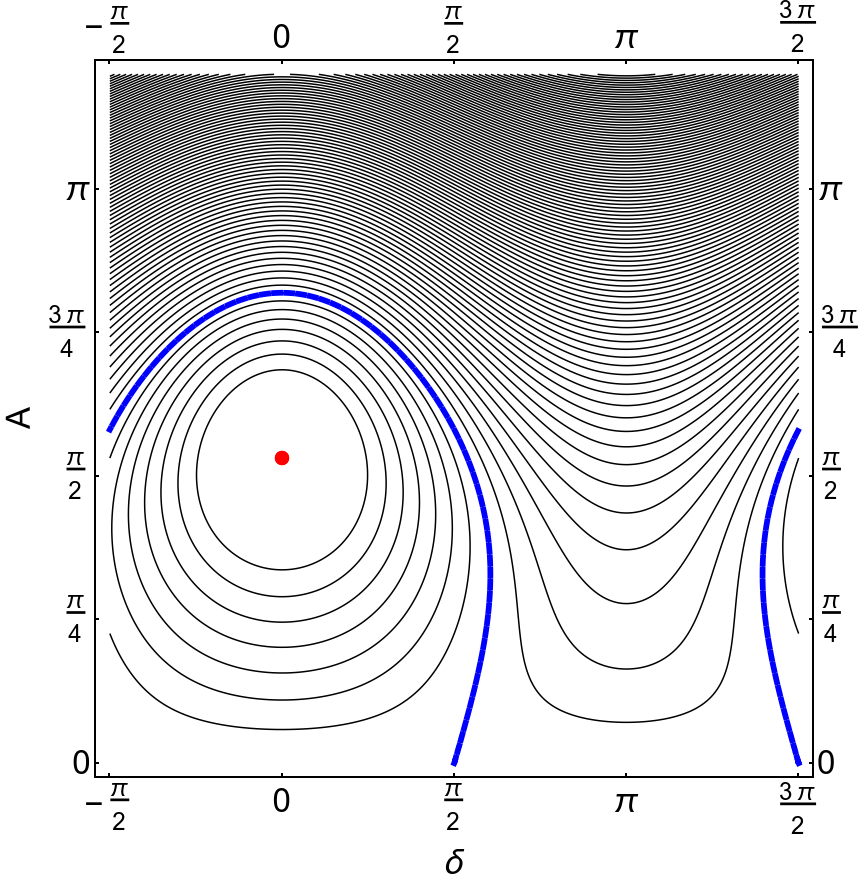} \quad b \includegraphics[scale=0.15]{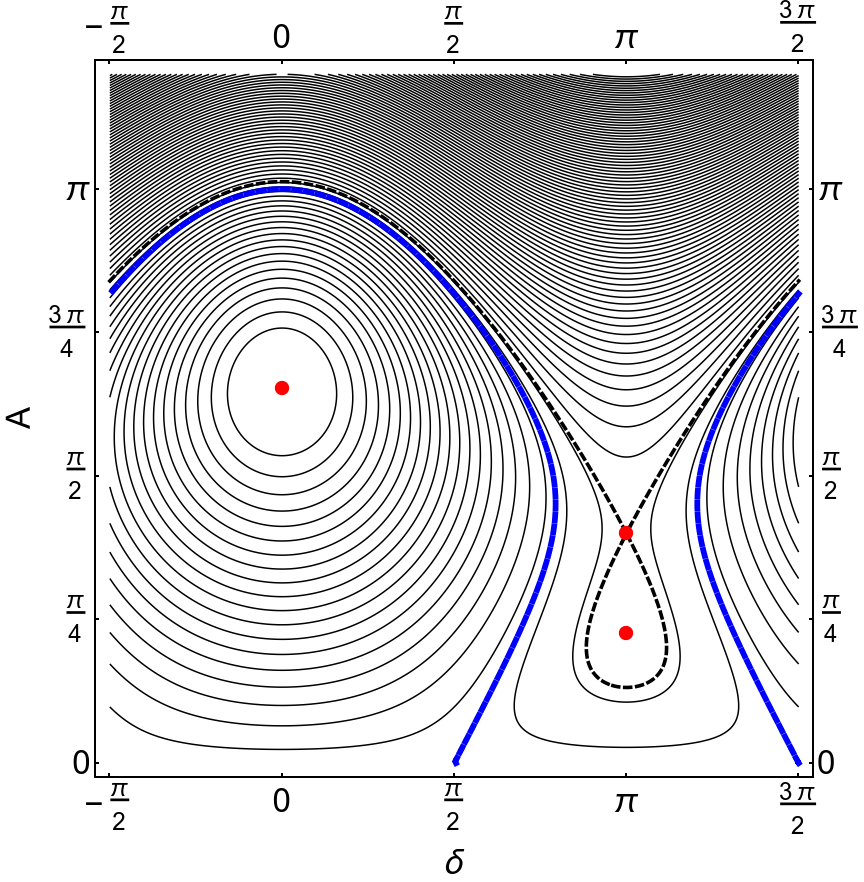} \\
 c \includegraphics[scale=0.15]{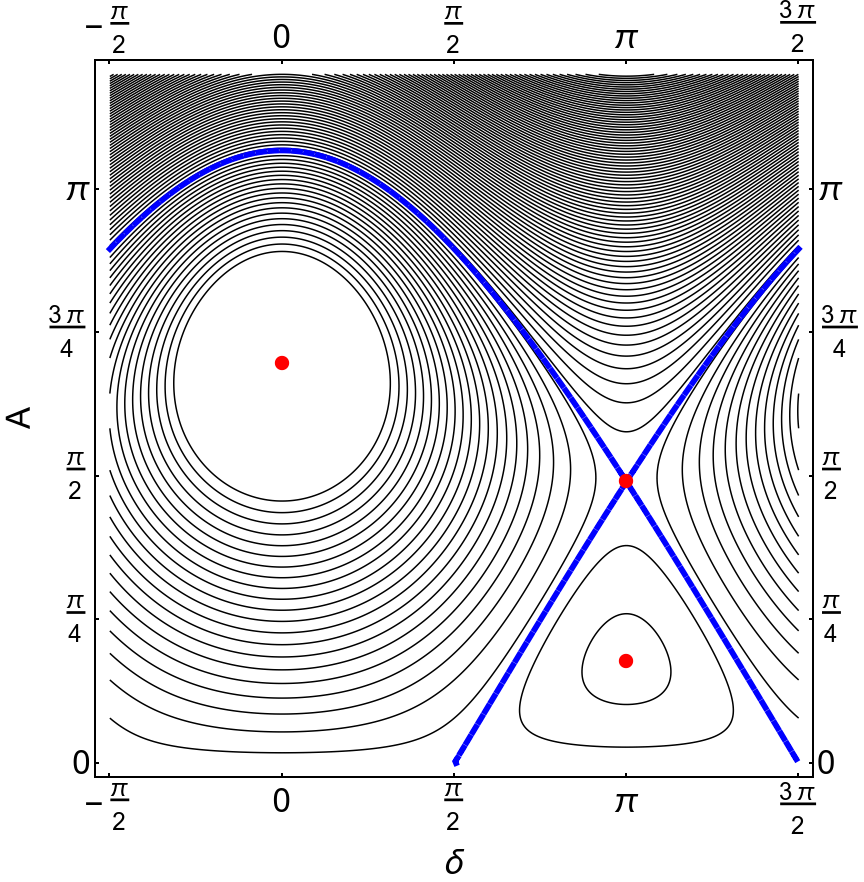} \quad d \includegraphics[scale=0.15]{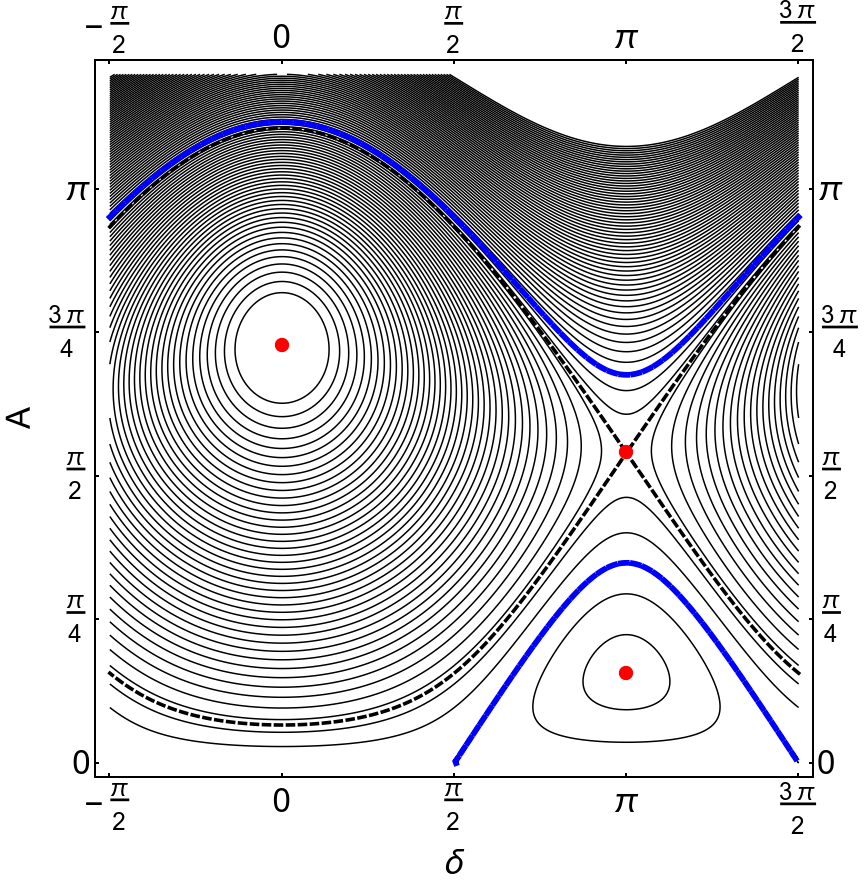} 
\caption{The phase portraits of  system (\ref{eq:Hforced}) at force amplitude $f=0.2$ and  different values of  frequency $\omega$. Panels (a-d) correspond to $\omega=0.9, 0.81, 0.776, 0.75$, respectively. The Limiting Phase Trajectories and separatrix are shown as thick blue and black dashed curves, respectively. Red points correspond to the stationary states.}
 }
 \label{fig:PPescape}
 \end{figure*}
Figure 3(a) represents the phase portrait with single stationary state at the phase $\delta=0$.
The thick blue curve, which passes through zero valued amplitude, separates the trajectories closed  around the stationary point from the transit-time ones.
This trajectory is called the Limiting Phase Trajectory (LPT).
In the problem under consideration the LPT describes the escape from the potential well, if the maximum of the LPT exceeds the limiting angle of the oscillations (i.e., $\pi$).
Such a case is observed in panel (b) of figure 3.
The value of the threshold frequency can be evaluated from the condition
\begin{eqnarray}\label{eq:omegaH}
H_f \left(a=0, \delta=0 \right) = H_f \left(a=\sqrt{\frac{\omega}{2}}  \pi, \delta=0 \right).
\end{eqnarray}
This condition corresponds to the high-frequency boundary of the escape from the well.
One should notice that the phase portrait on panel (b) contains three stationary points, therefore, the respective frequency is smaller then frequency $\omega_*$ mentioned above.
Decreasing  frequency $\omega$ we can obtain the phase portrait depicted on panel (c) of figure 3.
The specific feature of this phase portrait is that the LPT coincides with the separatrix crossed the unstable stationary point at $\delta = \pi$.
No possibility to escape the potential well at this frequency and at smaller ones (see fig. 3(d)) occurs because trajectory, which starts at $a=0$ can not reach the limiting angle $\pi$.
The frequency corresponding to the phase portrait on panel (c) can be evaluated by solving of the equation
\begin{eqnarray}\label{eq:omegaL}
H_f(a=0, \delta=0) = H_f ( a=a_u, \delta = \pi ),
\end{eqnarray}
where $a_u$ corresponds to the unstable stationary point.

Thus solving equations (\ref{eq:omegaH}) and (\ref{eq:omegaL}), we can determine the domain of the force's frequency and amplitudes, where the escape from the potential well is possible.
It is noteworthy that, in order to do it, we do not need in  solving the non-stationary equations (\ref{eq:adelta0}), but we can find the domain's boundaries analysing the variation of the phase portrait at various values of $f$ and $\omega$.

Nevertheless, if we want to estimate the time of the escape from the well, we need in the integration of the non-stationary equation along the LPT.
Actually, the escape time can be estimated as follows:

\begin{eqnarray}\label{eq:Tescape}
T= \int_{0}^{T}{dt}=\int_{0}^{\sqrt{\frac{\omega}{2}}\pi}{\frac{da}{da/dt}}
\end{eqnarray}
From the first of equations (\ref{eq:adelta0}) we get
\begin{eqnarray}\label{eq:dadt}
\frac{d a}{d t}=\frac{f \sin{\delta}}{2\sqrt{2\omega}}
\end{eqnarray}
Taking into account that the LPT passes through zero amplitude, the expression for $\cos{\delta}$ can be found from energy (\ref{eq:Hforced}):
\begin{eqnarray}\label{eq:cosd}
\cos{\delta}=\frac{\sqrt{2\omega} }{a f}\left(1- \frac{\omega}{2} a^2- J_0\left(\frac{\sqrt{2} a}{\sqrt{\omega }}\right)\right)
\end{eqnarray}
Finally, combining equations (\ref{eq:Tescape} -- \ref{eq:cosd}), we can write the period of the escape from the well as follows:
\begin{widetext}
\begin{eqnarray}\label{eq:Tescape2}
T=\int_0^\pi{\frac{ \sqrt{2 \omega} A  dA }{\sqrt{\left(- \left(\frac{\omega}{2} A \right) ^2- J_0(A)+ \frac{ f}{2} A+1 \right) \left(\left(\frac{\omega}{2} A \right) ^2+ J_0(A)+\frac{  f}{2} A-1 \right)} }}
\end{eqnarray}
\end{widetext}
The last expression can be estimated numerically at fixed values $\omega$ and $f$.
Figure 4 shows the contours of the constant escape time, which have been calculated accordingly to expression (\ref{eq:Tescape2}).
The uncoloured region corresponds to ($\omega-f$) domain, where the transition to rotation of pendulum is unreachable.
\begin{figure}
\centering{
\includegraphics[scale=0.2]{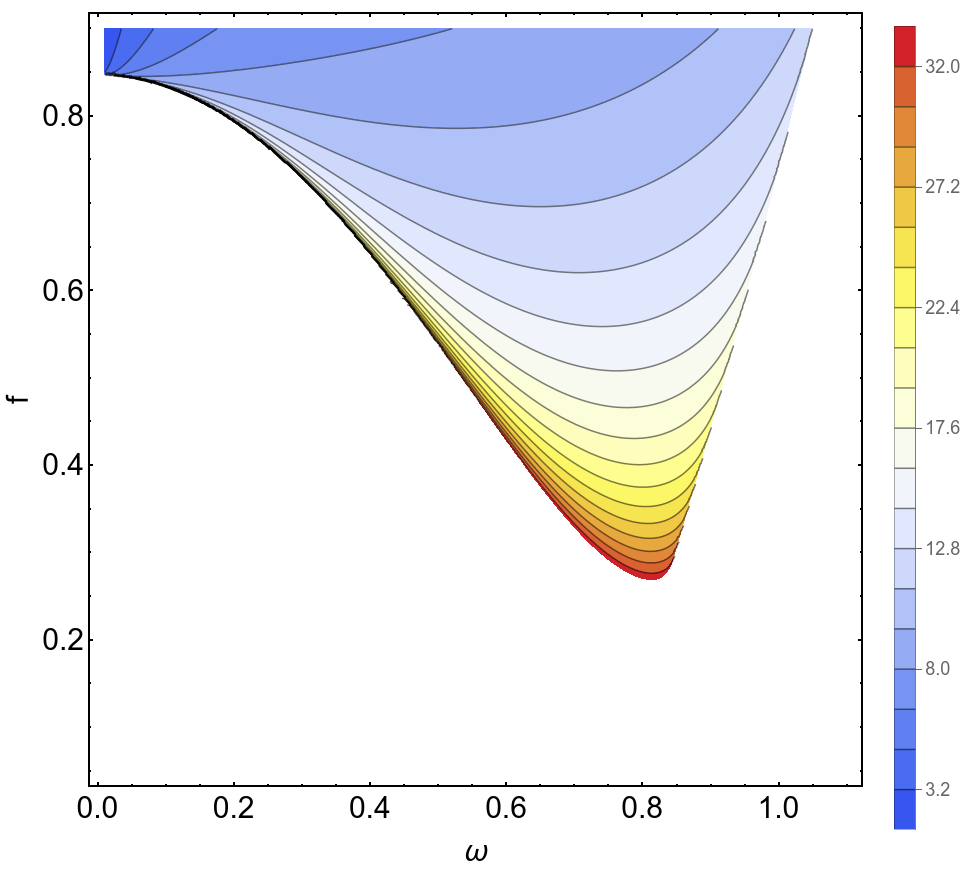}
\caption{The time of the escape from potential well on the $\omega - f$ plane. Contours correspond to the constant periods of the escape from the well, which are signed in the plot legend in right.}
}
\label{fig:Tescape}
\end{figure}

\subsection{Instability of the rotation of coupled pendula}\label{Rotation}
In this section we would like to perform the stability analysis for the in-phase rotation of two coupled pendula.
More detail description should be looked  over \cite{Smirnov2019}.

The energy of two coupled pendula  is determined as follows:
\begin{widetext}
\begin{equation}\label{eq_Hdouble}
H=\sum_{j=1,2} {\left( \frac{1}{2} \left(\frac{d q_j }{d t}\right)^2+\sigma  (1-\cos {q_j})+\frac{\beta}{2} \left(1-\cos{\left(q_j-q_{3-j}\right)} \right) \right)}.
\end{equation}
\end{widetext}
(In order to allow the  mutual rotation of the pendula, we assume $2 \pi-$ periodical interpendulum potential, which is similar to the interaction of the coaxial arranged dipoles.)

The equations of motion have the form
\begin{equation}\label{eq_EMdouble}
\frac{d^2 q_j }{d t^2}-\beta \sin{\left(q_{3-j}-q_j \right)}+\sigma  \sin {q_j }=0; \quad j=1,2.
\end{equation}

Let us introduce the in-phase mode $\theta_1=(q_1+q_2)/2$ and the out-of phase mode $\theta_2=(q_1-q_2)/2$.
The respective equations of motion

\begin{eqnarray}\label{eq_eqmodes}
\frac{d^2 \theta_1}{d t^2} + \sigma \cos{\theta_2}\sin{\theta_1}=0, \\ \nonumber
\frac{d^2 \theta_2}{d t^2}+ \beta \sin{2 \theta_2} + \sigma \cos{\theta_1} \sin{\theta_2}=0.
\end{eqnarray}
admit the exact solutions $\left( \theta_1=\theta_1(t), \theta_2=0 \right)$ and $\left( \theta_1=0, \theta_2=\theta_2(t) \right)$.

If  energy $E$  of in-phase motion exceeds the value $2 \sigma$, the pendula undergo the synchronous rotation with period
\begin{eqnarray}\label{eq:period}
T=2 \sqrt{2} \frac{K \left(\frac{2 \sigma}{E} \right)}{\sqrt{E}},
\end{eqnarray}
where $K$ is the complete elliptic integral of the first kind.
However, the numerical simulations show that the in-phase rotation turn out to be unstable at some values of coupling parameter $\beta$.
The difference of the rotation velocities of the pendula exhibits some periodic perturbations, an example of which is shown in figure 5(a).
\begin{figure*}
\centering{
a \includegraphics[scale=0.2]{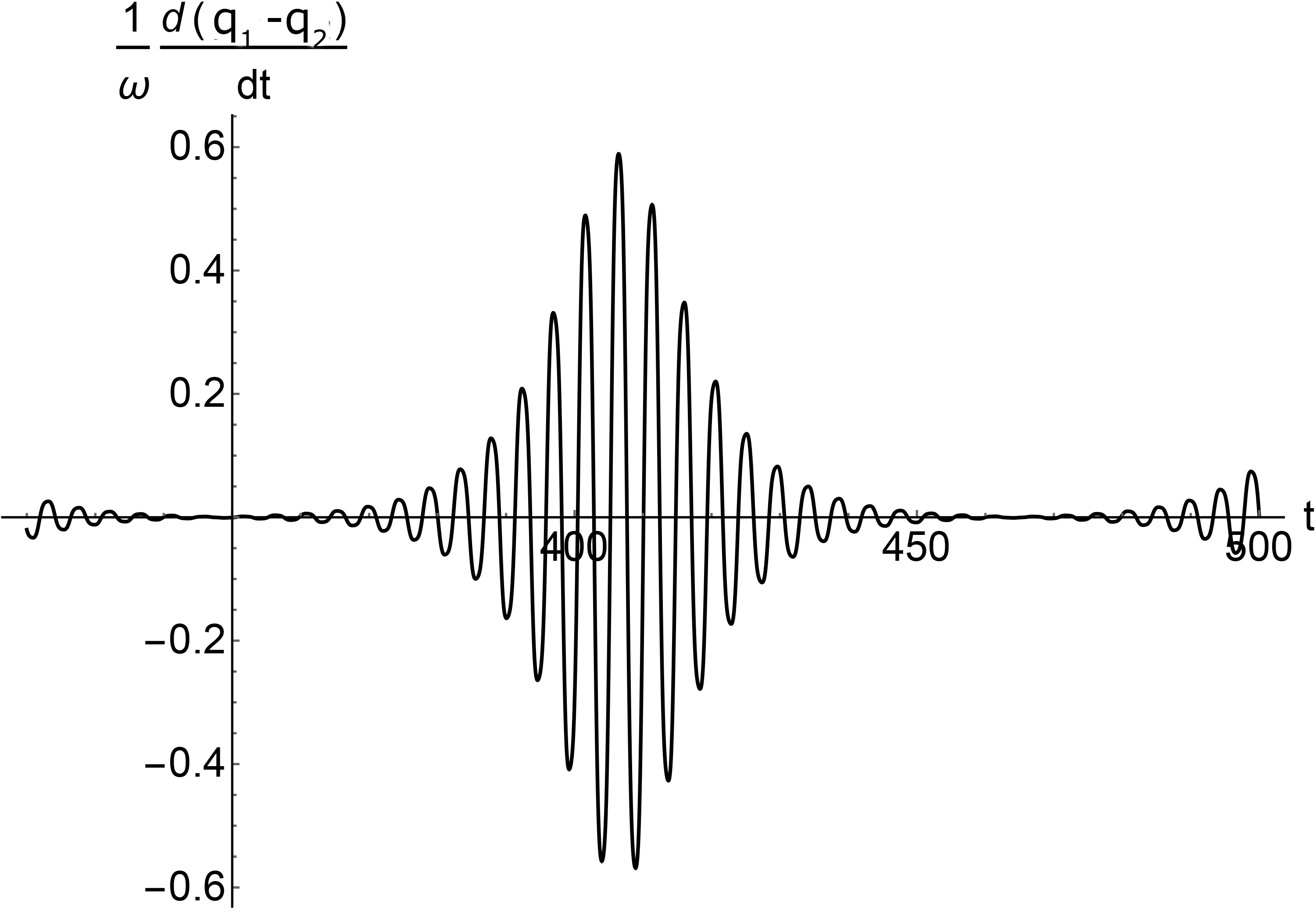} \quad  b \includegraphics[scale=0.2]{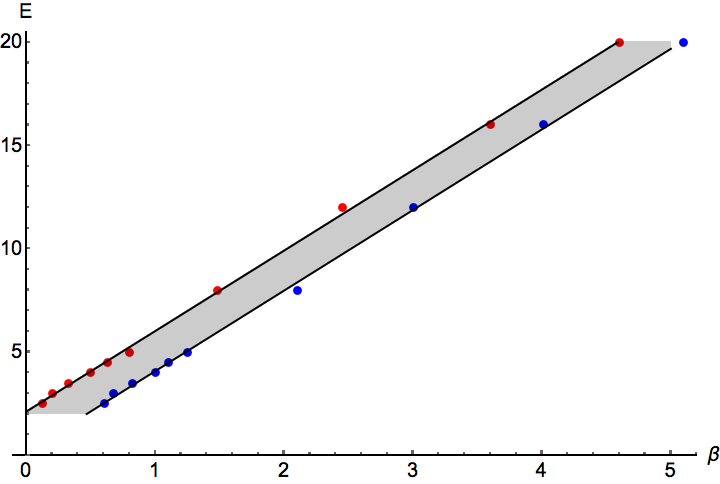}
\caption{(a) The data of numerical simulations of the in-phase rotation of coupled pendula.  The behaviour of the difference of the pendulums velocities in the unstable range of the in-phase rotation. (b) The range of the coupling constant $\beta(E)$, where the unstable in-phase rotation occurs, is coloured gray. Blue and red points were obtained in the direct numerical simulations.  The black lines show the threshold values, corresponding to relations (\ref{eq_threshold1}, \ref{eq_threshold2}).}
}
\label{fig:instability0}
\end{figure*}
The panel (b) of figure 5 shows the range of  coupling parameter $\beta$ with unstable rotations of the pendula in dependence on the rotation energy.
At the first glance the existence of instability in the in-phase rotation contradicts to the limiting cases of the extremely large and extremely small coupling.
Actually, if the coupling is negligible, the pendula are independent and no instability in the rotation occurs.
From the other end, when coupling constant $\beta \rightarrow \infty$, two pendula can be considered as a single pendulum with the doubled mass and no instability occurs again.

Let us assume that the energy of the in-phase rotation is large enough $(E \gg 2\sigma)$. 
In such a case we can represent the obvious solution of the first of equation (\ref{eq_eqmodes}) as follows
\begin{eqnarray}\label{eq:approximate0}
\theta_1=\omega_r t + \lambda \sin{\omega t},
\end{eqnarray}
where $\omega_r=2 \pi / T$ is the rotation frequency and $\lambda =\sigma / \omega_r^2 \ll 1$.

In order to analyse the stability of the in-phase rotation, we should consider the out-of-phase variable as a small perturbation.
Taking into account solution (\ref{eq:approximate0}), we can write
\begin{equation}\label{eq_psi2_1}
\frac{d^2 \theta_2}{d t^2}+\left(2\beta +\sigma \cos{\omega_r t} \right) \sin{\theta_2} = 0.
\end{equation}
(Considering $\theta_2$ as a small perturbation we assume that $\cos{\theta_2} \approx 1$.)

Equation (\ref{eq_psi2_1}) is  the well-known equation of the parametrically excited pendulum.
It is common knowledge that the first parametric resonance occurs at the frequency, which is one-half of the own one.
Thus we should estimate the perturbation with frequency $\Omega= \omega_r /2$.

Representing out-of-phase perturbation in the complex form accordingly expression (\ref{eq:Phi0}), extracting the "carrier" exponential $ e^{- i \Omega t}$ and discriminating the secular term, we can write the stationary equations for the modulus and the phase of the perturbation:
\begin{widetext}
\begin{eqnarray}\label{eq_eqparam}
&\frac{\sigma}{a} J_2 \left( \sqrt{\frac{2}{\Omega}} a \right) \sin{2 \delta} =0 \\ \nonumber
&\frac{\Omega}{2} a -\frac{1}{\sqrt{2\Omega}}  2 \beta J_1 \left( \sqrt{\frac{2}{\Omega}} a \right)   
  -\frac{\sigma}{\sqrt{2\Omega}} \left[ J_1 \left(\sqrt{\frac{2}{\Omega}} a \right) - \frac{\sqrt{2 \Omega}}{a} J_2 \left(\sqrt{\frac{2}{\Omega}} a \right)  \right]  \cos{2 \delta}  =0,
\end{eqnarray}
\end{widetext}
where $J_n$ is the Bessel function of order $n$.
These equations describe the stationary out-of-phase oscillations coupled with the in-phase rotation of the pendula.
The amplitude and phase of such oscillations should be determined numerically.
However, we can write the energy of such oscillations as follows

\begin{equation}\label{eq_Hparam}
H=\frac{\Omega}{2} a^2 - 2 \beta \left( 1-J_0 \left(\sqrt{\frac{2}{\Omega}} a \right) \right) - \sigma J_2 \left(\sqrt{\frac{2}{\Omega}} a \right) cos{2 \delta}.
\end{equation}
Taking into account this expression, one should analyse the non-stationary trajectories on the phase plane at the different values of coupling parameter $\beta$.

Figure 6 shows the phase portraits of the system with rotation energy $E=5.0$ at three specific values of $\beta$.
Fig. 6(a) represents the phase plane of the system when the coupling parameter $\beta$ is smaller than the bottom threshold of the instability. 
One can see that there is only stationary solution $a=0$ for any values of phase $\delta$.
Any trajectories, which are close to the stationary state can not rise and the rotation is stable.

\begin{figure}
\centering{
a\includegraphics[scale=0.125]{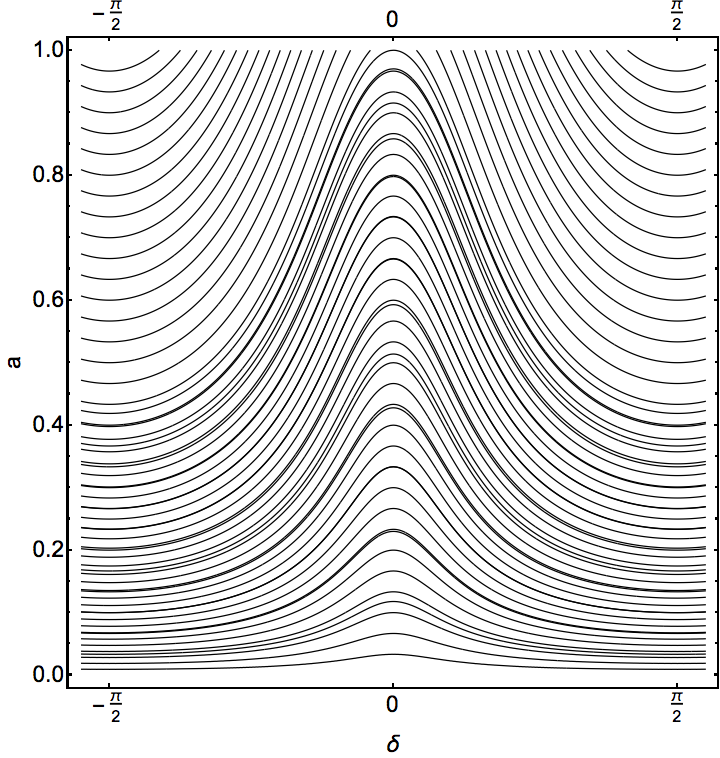}  b\includegraphics[scale=0.125]{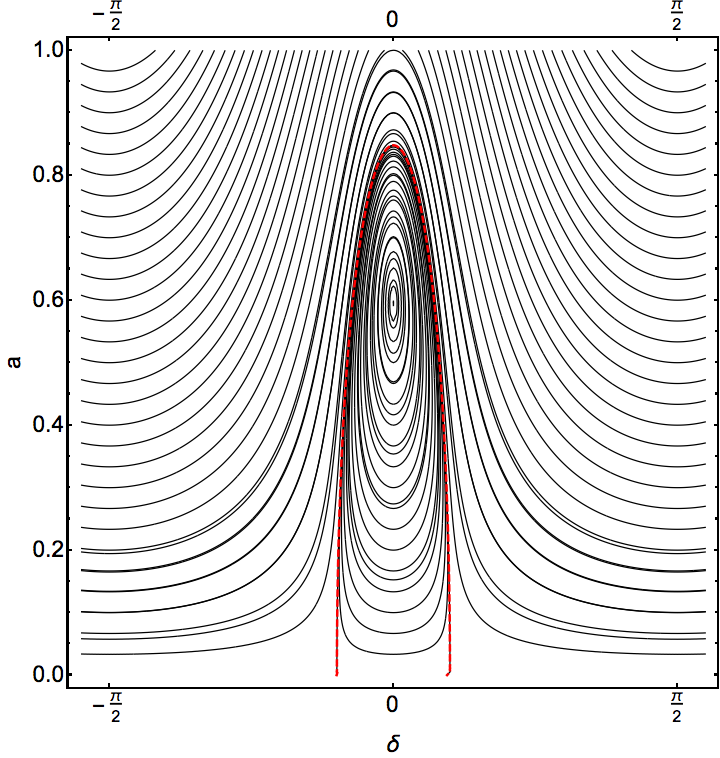} 
c\includegraphics[scale=0.125]{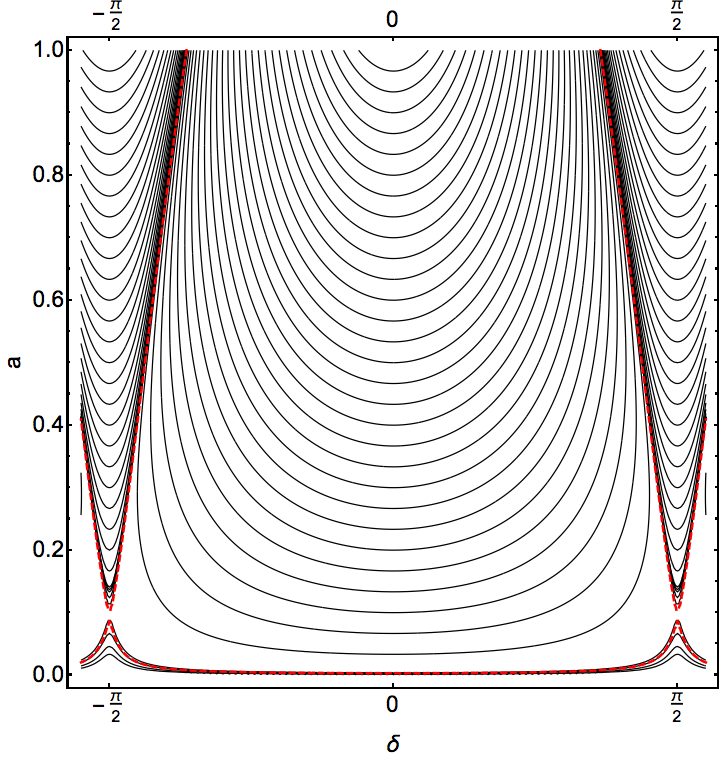}  
\caption{(Color online) The phase plane of system (\ref{eq_Hparam}) at three values of coupling parameter $\beta$ (a) $\beta =  0.7$, (b) $\beta =0.9$, (c) $\beta=1.2275$. $E=5.0$, $\sigma=1$.}
}
\label{fig:5_1}
\end{figure}

However, if coupling constant $\beta$ exceeds some threshold, new stationary point with $a\neq 0$ and  phase $\delta = 0$ appears.
Simultaneously the trajectory, which separates the sets of the closed and transit-time trajectories, forms.
The fact is important that this trajectory pass through zero value of the amplitude $a$, therefore any perturbations, which start from zero amplitude evolve along this trajectory.
This trajectory is the aforementioned LPT.
Fig. 6(b) shows the phase portrait after formation of the stationary solution and the LPT.
One should notice that from the viewpoint of the rotational instability,  it is not the fact of the existence of the stationary point that matters, but the appearance of the Limiting Phase Trajectory.

In order to determine the threshold of the instability, one should note that the latter arises via the  formation of the stationary solution at the point $(a=0, \delta=0)$.
This needs in the  condition $\partial^2 H_2/\partial a^2 =0$ at this point.
Solving this  equation with respect to the coupling parameter, we get the bottom threshold value as follows

\begin{equation}\label{eq_threshold1}
\beta_{l}=\frac{1}{4} \left( 2 \Omega^2- \sigma \right).
\end{equation}

While the coupling parameter grows the amplitude of the stationary solution as well as the LPT  increase too.
However, when the LPT's low edges reach the phase value $\delta= \pm \pi /2$, the next bifurcation happens.
Namely, the symmetrical pair of the unstable stationary solutions appear in the points $(a=0, \delta = \pm \pi /2)$.
Simultaneously, the LPT transforms to the separatrix crossing  these unstable states.
Figure 6(c) shows the phase plane after second bifurcation.
Because of the same branch of the separatrix passes through both unstable points,  no small-amplitude trajectories can grow.
The value of the coupling parameter corresponding to this bifurcation can be estimated from the criterion of stationary point creation with phase $\delta = \pm \pi/2$, that leads to 
\begin{equation}\label{eq_threshold2}
\beta_{t}=\frac{1}{4} \left( 2 \Omega^2 + \sigma \right).
\end{equation}
The thresholds  (\ref{eq_threshold1}, \ref{eq_threshold2}) are shown in figure 5(b) by the black lines.
One can notice the excellent fit these values with the data of the numerical simulations.

Three problems discussed above exhaust the list of the examples, by which we would like to illustrate the applications of the CEVA, but many another applications  of the method discussed above can be found in  \cite{Manevitch2018}.

\section{Discussion}\label{Conclusion}

In spite of the list of the possible applications of the CEVA can be continued we would like to discuss the method's peculiarities. 
First of all, one should notice that the CEVA is close to such widely used methods as the Van der Pol one, the multiscale expansion, and the harmonic balance approximation. 
Like two first methods the CEVA uses the envelope functions for the description of the stationary as well as non-stationary dynamics. 
Nevertheless, due to   application of the complex variables the final results are more clear and understandable. 
Moreover, in the number of cases we can get the frequency spectrum (or the frequency-amplitude relation) without tedious calculations for large amplitude oscillations (see, for example, Appendixes), and we can do it even in the presence of the external forces and the friction. 
In fact, no assumption about the smallness of amplitude or the presence of any small parameter is needed. 
The natural restriction for this procedure arises from the requirement of the sufficient smoothness of the the frequency as the function of the oscillation amplitude. 
In other words, the single-frequency approximation has to be sufficient and the influence of higher harmonics should be neglegible. 
Really, the accounting the third harmonic in the framework of the CEVA is possible \cite{Lamarque2020}, but  above we restricted by the single-harmonic description. 
The natural sequence of the single-frequency description is that the CEVA is the insufficient in the neighbourhood of separtrix solutions.
The disadvantage of the CEVA is that the using the stationary solution in the form (\ref{eq:Psistat}) and averaging the equations (\ref{eq:complexfull}) lead to that only symmetric part of the potential makes the contribution into oscillation frequency. 
Solving this problem is the subject of further researches.

While the stationary equations are clear enough, the development of the non-stationary  equations in the framework of the CEVA need in additional accuracy. 
The reason is that the "slowness" of the non-stationary equations depends on a smallness of the right hand side of equations (\ref{eq:nonstateqn},\ref{eq:adelta0}). 
At the same time, the amplitude of the disturbed solution can be essentially different than the amplitude of the stationary oscillation, but the frequency should be approximately the same. 
By what manner we can satisfy so opposite requirements? 
As an example we should  point the resonant interaction of the nonlinear normal modes. 
Actually, the frequencies of modes with close wave number are close (see Appendix A) if their amplitudes are equal. 
However, amplitude of their sum  in dependence of the phase difference can be almost redouble or turns out to be small. 
Due to smallness of the frequency difference the transition from one state to another one describes by the slow non-stationary dynamical equations. 
(This process is an analogue of the well known beating in the system of  two weakly coupled identical oscillators.)
Thus, we need in the careful control of the time scale of the non-stationary equations.
 Another problem with the non-stationary equations, which are deduced accordingly to relations (\ref{eq:nonstateqn},\ref{eq:adelta0}), associates with the complexity of them.
 They admit the analytical solution only for the exceptional cases even for the simplest power potentials.
However, some simplification originates from the existence of the additional integral of motion, which is analogue of the quantum-mechanical occupation number. 
For example, in the case of N-particle lattice this integral is $X=\sum{|\psi_j|^2}$ \cite{Smirnov2010}. 
In the number of cases the presence of this integral allows us to reduce the dimensionality of the phase space and to do the analysis by the phase plane method.

One should emphasize that the using of  Hamilton function (\ref{eq:H1}) in the space of the envelopes turns out to be very successful, because it allows us to make some principal findings without solving the non-stationary equations.
It was demonstrated in the analysis of the in-phase rotation instability of the coupled pendula.
The analogous results were obtained in \cite{SmirnovLA2016} by essentially large efforts.
In this work we do not discuss the analysis of the interaction of the nonlinear normal modes,  because of this problem has been described in the number of papers for the discrete dynamical systems of various origin (coupled oscillators, coupled self-excited oscillators, nonlinear chains, carbon nanotubes).
Only one remark should be made about transition to the infinite degrees-of-freedom systems.
It was demonstrated that the continualization of the non-stationary equations for the  discrete systems leads to the specific nonlinear Schr\"{o}dinger equation.
The latter turns out to be the complex analogue of the well-known sine-Gordon equation in the case of the Frenkel-Kontorova or the Sine-lattice models \cite{Smirnov2017,Manevitch2018} , or describes the specific breather solution for the circumferential flexure oscillations of the carbon nanotubes \cite{Smirnov2016PhysD,Manevitch2018}.

\appendix   

\section{The dispersion relation for 1D Sine-lattice}

In the Appendix we clarify the derivation of the stationary equation in terms of complex variables.
We will use the chain of the interacting particles with $2 \pi$-periodic potential (so called sine-lattice \cite{HommaTakeno1984,Takeno1986,Smirnov2017}).
The Hamilton function of the chain can be written as follows:
\begin{widetext}
\begin{eqnarray}
H= \sum_{j=1}^N{ \left( \frac{1}{2} \left(\frac{d \varphi_j}{d t} \right)^2 + \beta \left(1-\cos{\left( \varphi_{j+1}-\varphi_{j} \right)} \right) + \sigma \left( 1- \cos{\varphi_j} \right) \right)},
\end{eqnarray}
\end{widetext}
where $\varphi_j$ is the displacement $j-th$ particle from ground state, $N$ is the number of the particles, $\beta$ and $\sigma$ are the constants.

The respective equations of motion are read as 
\begin{eqnarray}
\frac{d^2 \varphi_j}{d t^2} - \beta \left( \sin{\left( \varphi_{j+1}-\varphi_{j} \right)} -\sin{\left( \varphi_{j}-\varphi_{j-1} \right)} \right) \\ \nonumber
+ \sigma \sin{ \varphi_j} = 0.  
\end{eqnarray}
Let us expand the trigonometric functions into Taylor series and replace the variables $\varphi_j$ accordingly to expressions (\ref{eq:inverse}):
\begin{widetext}
\begin{eqnarray}
i \frac{d \Psi_j}{d t} - \frac{\omega}{2} \left( \Psi_j - \Psi_j^* \right) + \frac{1}{\sqrt{2 \omega}} \sum_{k=0}^{\infty} \frac{(-1)^k}{(2k+1)!} \left(\frac{1}{\sqrt{2\omega}} \right)^{2k+1} \times  \\ \nonumber  
\left( \beta \left( \left(\Psi_{j+1}-\Psi_j + cc \right)^{2k+1}-\left(\Psi_{j}-\Psi_{j-1} + cc \right)^{2k+1} \right) -\sigma \left( \Psi_j+\Psi_j^* \right)^{2k+1} \right)  =0
\end{eqnarray}
\end{widetext}
In order to find the stationary single-frequency solution one should represent functions $\Psi_j \left( t \right) = \psi_j e^{- i \omega t}$ with  modulus $\psi_j$ which does not depend on the time.
It is easy to show that substituting this expression into equation (A3) and averaging it over period $2 \pi/\omega$ leads to the equation
\begin{widetext}
\begin{eqnarray}
\frac{\omega}{2} \psi_j + \frac{\beta}{\sqrt{2 \omega}} \Biggl( J_1 \left(\sqrt{\frac{2}{\omega}} |\psi_{j+1}-\psi_j| \right) \frac{\psi_{j+1}-\psi_j}{|\psi_{j+1}-\psi_j |} -     \\  \nonumber
J_1 \left(\sqrt{\frac{2}{\omega}} |\psi_{j}-\psi_{j-1}| \right) \frac{\psi_{j}-\psi_{j-1}}{|\psi_{j}-\psi_{j-1} |} \Biggr) - \frac{\sigma}{\sqrt{2 \omega}} J_1 \left(\sqrt{\frac{2}{\omega}} |\psi_j| \right) \frac{\psi_j}{|\psi_j |} =0,
\end{eqnarray}
\end{widetext}
where  $J_1$ is the Bessel function of the first order.
Assuming the periodic boundary conditions and taking into account relation (\ref{eq:Ypsi}) one can see that the plane wave $\psi_j = \chi e^{i \kappa j}$ with $\kappa = 2 \pi k/N$, $k=0,1,\dots, N/2$ and $\chi=\sqrt{\omega/2} A$ is the exact solution of equation (A4), if frequency $\omega$ is determined by expression
\begin{eqnarray}
\omega^2=\frac{2}{A} \left( \sigma J_1\left( A \right) + 2 \beta J_1 \left( 2 A \sin{\frac{\kappa}{2}} \right) \sin{\frac{\kappa}{2}} \right)   
\end{eqnarray}

\section{Estimation of the frequencies of simple non-linear oscillator }

In order to demonstrate the efficiency of the CEVA in the estimation of the amplitude-frequency relation for the  stationary oscillations, we made the calculation for the simple one degree-of-freedom system:

\begin{eqnarray}
\frac{d^2 x}{d t^2}+x+x^3+x^5 = 0 %
\end{eqnarray}

Following the procedure of section 2, it is easy to show that the frequency of the stationary oscillations is determined as follows:

\begin{eqnarray}
\omega_1=\sqrt{1+\frac{3}{4} A^2+\frac{5}{8} A^4} 
\end{eqnarray}
These results we compare with the data, which have been obtained in works \cite{He2019} and \cite{Chowdhury2017} by different methods.
The final expression for the frequency in work \cite{He2019} was obtained by some intuitive considerations:

\begin{eqnarray}
\omega_2=\sqrt{1+0.83 A^2+0.51783 A^4} 
\end{eqnarray}
In spite of that this expression is extremely similar to equation (B2), we suppose that the choice of the numerical coefficients is rather intuitive \cite{He2019}.
The respective values of the oscillation frequency are shown in Table 1 as $\omega_2$. 

The method of high-order harmonic balance was used in work \cite{Chowdhury2017}.
After some tedious calculations  the values of the oscillation frequency for different amplitudes have been obtained.
They are represented in Table 1 as  $\omega_3$. 
%

The main problem of the last two work is that even a little changing  the initial equation (B1) (for example, a varying of the numerical coefficients)  can cause  arduously predictable effect on the final result and, in order to estimate it, we should perform  the full calculation more times.
Table 1 and figure 7 represent  the comparative analysis of the data, which were calculated by expressions (B2) and reprinted from works \cite{He2019} and \cite{Chowdhury2017}.
The relative errors have been calculated with respect to  exact values ($\omega_e$) (see Table 1). 
\begin{eqnarray}
\delta_j=\frac{\omega_j-\omega_e}{\omega_e}, (j=1,2,3)  
\end{eqnarray}

\begin{table}[h!]
\caption{The frequencies of the stationary oscillations for  system (B1) at different amplitude $A$. $\omega_e$ corresponds to the exact value, $\omega_1$ is determined by expressions (B2), and $\omega_2$, and $\omega_3$ reprinted from works \cite{He2019} and \cite{Chowdhury2017}, respectively. Relative errors $\delta_j$ are calculated accordingly to expression (B4).}
\label{tab:1}       
\begin{ruledtabular}
\begin{tabular}{cccccccc}
$A$ & $\omega_e$ & $\omega_1$ & $\omega_2$ & $\omega_3$ & $\delta_1$ & $\delta_2$ & $\delta_3$
\\ 
\noalign{\smallskip}\hline\noalign{\smallskip}
 0.1 & 1.0038 & 1.0038 & 1.0042 & 1.0038 & -0.0004 & 0.0396 & 0.0003 \\
0.3 & 1.0356 & 1.0357 & 1.0387 & 1.0358 &-0.0102 & 0.3051 & 0.0226 \\
0.5 & 1.1065 &1.1075 &1.1135 & 1.1084 & -0.0870 & 0.6283 & 0.1635 \\
1.0 & 1.5236 & 1.5411 & 1.3718 & 1.5485 & -1.149 & 9.9639 & 1.6332 \\
 3.0 & 7.2686 & 7.6404 & 7.1003 & 7.7265 & -5.114 & 2.3156 & 6.2990 \\
 5.0 & 19.1815 & 20.2577 & 18.584 & 20.5097 & -5.611 & 3.1129 & 6.9244 \\
\end{tabular}
\end{ruledtabular}
\end{table}

\begin{figure}[h]
\centering{
\includegraphics[scale=0.2]{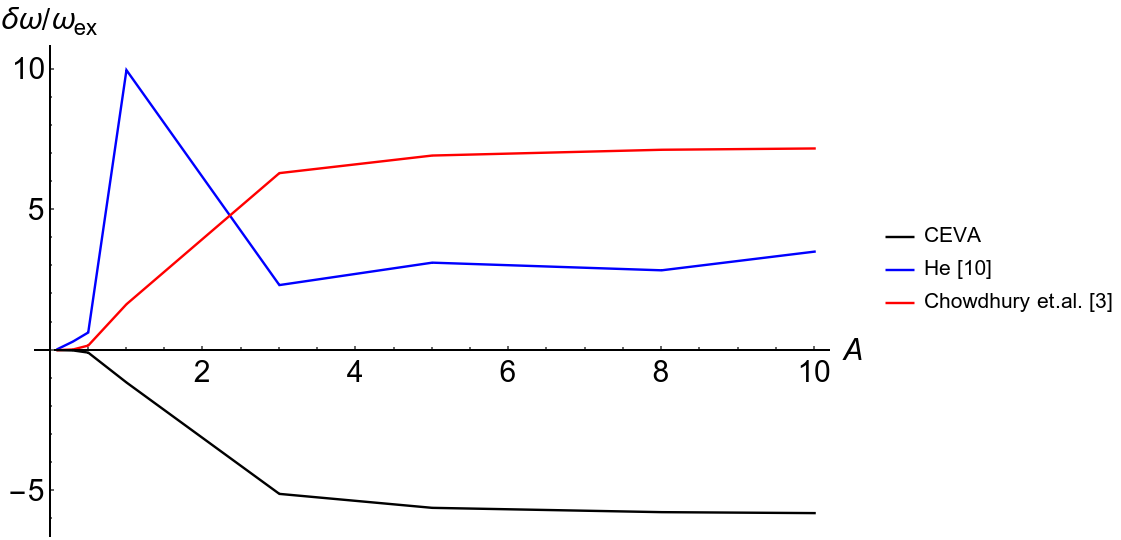}
\caption{The relative errors for the oscillation frequencies. Black, blue and red lines show the results of current work, and the data from work \cite{He2019} and \cite{Chowdhury2017}, respectively. }
}
\label{fig_ApB}
\end{figure}

\begin{acknowledgments}
Authors are grateful to Russia Science Foundation (grant 16-13-10302) for the financial supporting of this work.
\end{acknowledgments}


\bibliography{CEVA}

\end{document}